\documentclass{emulateapj}
\usepackage{natbib}
\usepackage{graphicx}
\usepackage{graphics,epsfig}
\usepackage{tabularx}
\usepackage{adjustbox,lipsum}
\usepackage{subfigure}
\usepackage{amsmath}
\usepackage{float}

\usepackage{color}
\usepackage{colortbl}
\usepackage[colorlinks=true,
                     linkcolor=red,
                     citecolor=blue]{hyperref}
\usepackage{courier}

\singlespace

\newcommand{\ba}{\begin{eqnarray}}
\newcommand{\ea}{\end{eqnarray}}

\newcommand  \beq    {\begin{equation}}

\newcommand  \eeq    {\end{equation}}
\newcommand  \gtsim  {\lower.5ex\hbox{$\; \buildrel > \over \sim \;$}} 
\newcommand  \ltsim  {\lower.5ex\hbox{$\; \buildrel < \over \sim \;$}}

\shortauthors{} 

\begin{document}

\title{Polarized galactic synchrotron and dust emission and their correlation}

\author{Steve K. Choi$^{1}$ and Lyman A. Page$^{1}$}
\email{khc@princeton.edu}
\altaffiltext{1}{Joseph Henry Laboratories of Physics, Jadwin Hall, Princeton University, Princeton, NJ 08544, U.S.A.}

\begin{abstract}
We present an analysis of the level of polarized dust and synchrotron emission using the WMAP9 and Planck data. The primary goal of this study is to inform the assessment of foreground contamination in the cosmic microwave background (CMB) measurements below $\ell\sim200$ from 23 to 353 GHz. We compute angular power spectra as a function of sky cut based on the Planck 353 GHz polarization maps. Our primary findings are the following. (1) There is a spatial correlation between the dust emission as measured by Planck at 353 GHz and the synchrotron emission as measured by WMAP at 23 GHz with $\rho\approx0.4$ or greater for $\ell<20$ and $f_{\mathrm{sky}}\geq0.5$, dropping to $\rho\approx0.2$ for $30<\ell<200$. (2) A simple foreground model with dust, synchrotron, and their correlation fits well to all possible cross spectra formed with the WMAP and Planck 353 GHz data given the current uncertainties. (3) In the 50$\%$ cleanest region of the polarized dust map, the ratio of synchrotron to dust amplitudes at 90 GHz for 50 $\leq \ell \leq$110 is $0.3_{-0.2}^{+0.3}$. Smaller regions of sky can be cleaner although the uncertainties in our knowledge of synchrotron emission are larger. A high-sensitivity measurement of synchrotron below 90 GHz will be important for understanding all the components of foreground emission near 90 GHz.

\end{abstract}


\section{Introduction}
\label{sec:intro}

The drive to characterize the CMB temperature and polarization anisotropy has led to a steady increase in our knowledge of foreground emission  from our galaxy. In this paper we focus on the diffuse polarized foreground emission in the 23-353 GHz regime as measured by the WMAP (\citeauthor{bennett/etal/2013}\citeyear{bennett/etal/2013}, hereafter B13) and the Planck \citep{planck_i/2015} satellites. There are two well established polarized components: synchrotron emission and thermal dust emission. It is known that they are spatially correlated as shown in \cite{kogut/etal/2007}, \cite{page/etal/2007}, and \cite{planck_xxii/2015} (hereafter K07, P07, and P15), that the polarized synchrotron is described by a power law with the spectral index varying over different galactic latitudes (K07; B13), and that the polarized thermal dust emission is well-described by a blackbody emitter with an emissivity that scales as a power law in frequency,
\begin{equation}
{\cal E}_d(\nu,\beta_d) \equiv \bigg(\frac{\nu}{\nu_{\rm{ref}}}\bigg)^{\beta_d - 2} \frac{B_{\nu}(T_{dust})}{B_{\nu_{\rm{ref}}}(T_{dust})},
\label{eq:dust_model}
\end{equation}
with $T_{dust}=19.6$~K and $\beta_d=1.59$ (P15), where $B_{\nu}(T)$ is the Planck function and $\nu_{\rm{ref}}=353$~GHz is the reference frequency.

We still have a lot to learn about foreground emission. For example, there is no definitive explanation for $\beta_d=1.59$ or for the observation that the spectral index of polarized synchrotron is different than that of unpolarized synchrotron \citep{gold/etal/2011}. It is also seen that the power in EE is roughly twice that in BB in both synchrotron and dust (P07; \citeauthor{gold/etal/2011} \citeyear{gold/etal/2011}; \citeauthor{planck_x/2015} \citeyear{planck_x/2015}; \citeauthor{planck_xxx/2014} \citeyear{planck_xxx/2014}, hereafter P14).  These observations are likely rooted in the configuration of the galactic magnetic field relative to the galactic plane and the coherence length of the magnetic field relative to the dust and electron distributions. Recently, for example, \cite{planck_xxxviii/2015} showed that the EE/BB ratio for dust may arise from the correlation between the filaments and the magnetic field orientations. Nevertheless, our characterization of foreground emission is largely empirical.

One of the notable findings is shown in Figure 16 of B13. The ``synchrotron spectrum" appears to flatten as one goes to higher frequencies and there is an apparent dust contribution. With synchrotron and dust parameterized as $T_s\propto\nu^{\beta_s}$ and $T_d\propto\nu^{\beta_d}$, a possible model is put forward that combines the $\beta_s$ flattening in frequency from $-3.13$ at 33 GHz to $-2.43$ at 94 GHz with an additional dust component that rises with $\beta_d=1.4-1.5$. The WMAP team cautiously interpreted this flattening as the synchrotron template tracing some of the dust due to shortcomings of the dust template. With the Planck 353 GHz polarization maps, we can now better test this model.

Our main goal is to test a simple model of polarized dust and synchrotron emission over different fractions of sky and angular scales. Often foreground studies are done with templates in map space. This procedure weights the largest angular scales because they have the highest signal to noise ratio. Here we examine the emission characteristics in $\ell$ ranges for masks of the sky relevant to searches for primordial B-mode polarization.

\section{Data}
\label{sec:data}

WMAP delivered all-sky polarization maps observed with ten independent differencing assemblies (DAs) covering five different frequency bands. We use the single year DA maps to examine the map statistics. We find this important for a thorough assessment of the angular power spectrum uncertainty. In particular, it gives us insight into the cross term between the signal and noise, and also allows us to investigate the distribution of the noise. Planck has released all sky polarization maps at 30, 44, and 70 GHz from the Low Frequency Instrument (LFI) and 100, 143, 217, and 353 GHz from the High Frequency Instrument (HFI). For this work, we use the full-mission and half-mission splits for only the 353 GHz maps. Intensity to polarization leakage in the 353 GHz maps, if not corrected, corresponds to $<4\%$ of the EE/BB power spectra for $\ell>50$ in the similar sky regions to what we define below (P14). We use the corrected maps. The precise level of the residual leakage at large angular scales is not yet published by the Planck collaboration.

\begin{figure}
\centering
\epsscale{1.2} 
\plotone{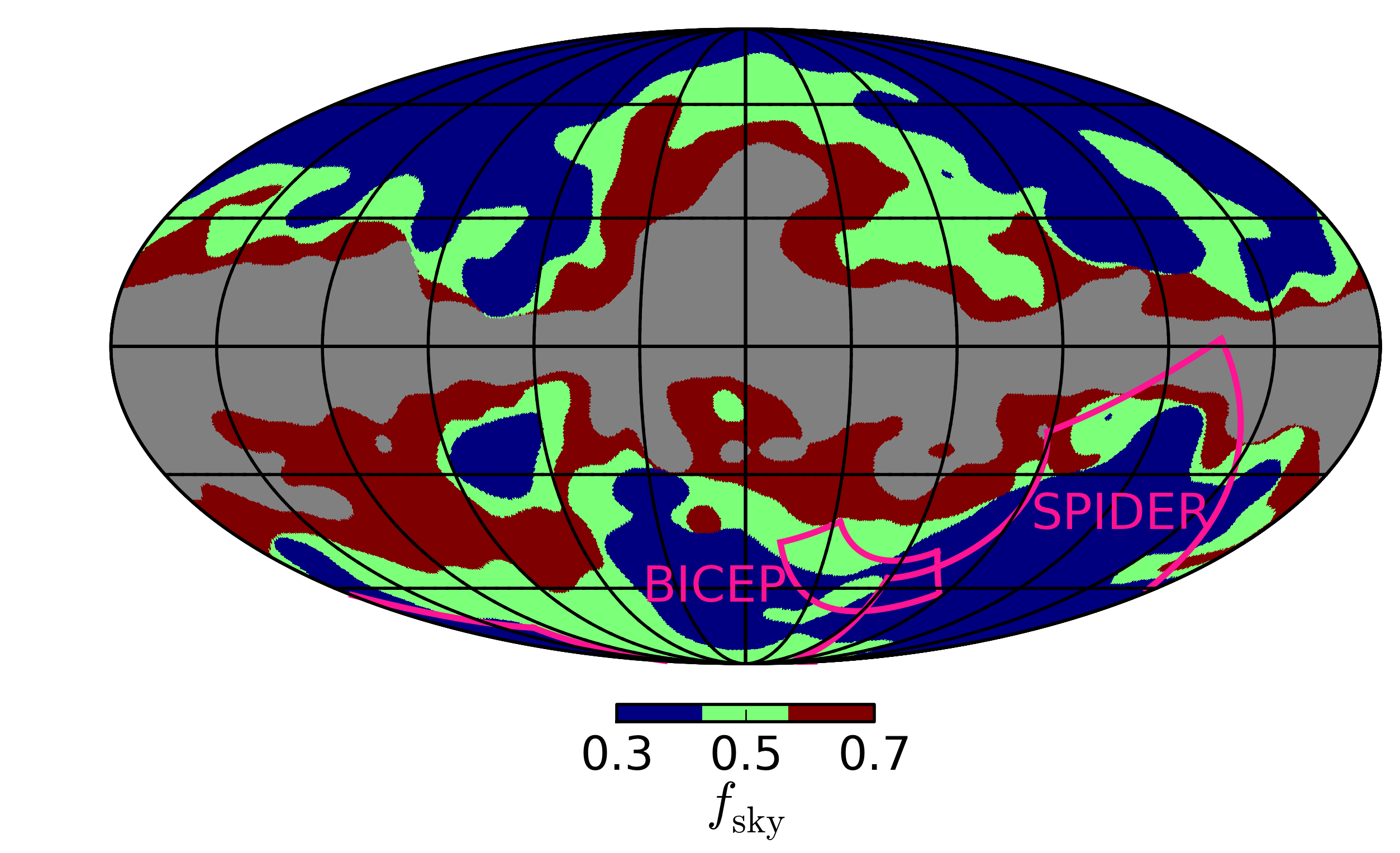}
\caption{Masks showing regions with incrementally larger levels of polarized emission in the Planck 353 GHz maps, with $f_{\rm{sky}}$ of 0.3, 0.5, and 0.7 (see details in Section~\ref{sec:data}). These regions do not necessarily indicate progressively higher levels of polarized emission for $50<\ell<110$. We also show the approximate regions observed by the BICEP2 and Keck experiments \citep{bicepkeck/2015} and SPIDER \citep{fraisse/etal/2013}.}
\label{fig:mask}
\end{figure}

We make masks with the Planck 353 GHz polarization map\footnote{We do not subtract the noise bias and the CMB from $P$ and this may have an effect on the exact definition of the regions.}, $P=\sqrt{Q^2+U^2}$, obtained from first degrading the Stokes $Q$ and $U$ maps to \texttt{HEALPix} \citep{gorski/etal/2005} $N_{\mathrm{side}}=128$ then smoothing them with a Gaussian FWHM of 1$^{\circ}$. We retain pixels above different thresholds set to get masks\footnote{We first make a binary mask by padding pixels over some threshold with $-1$ and the rest with $1$, then smooth the mask, set the negative pixels to 0 and retain the positive pixels to form largely contiguous masks with $f_{\rm{sky}}$ of 0.7, 0.5, and 0.3.} with $f_{\rm{sky}}$ of 0.7, 0.5, and 0.3 as shown in Figure~\ref{fig:mask}. The regions defined this way sequentially mask more of the galactic plane and high polarized emission regions. This approach selects the regions with incremental levels of power in the large angular scales, as these modes have the highest signal to noise ratio. We apodize the three masks with a Gaussian FWHM of 5$^{\circ}$. Since we are interested in analyzing the diffuse foreground emission, we mask all sources found in the WMAP point source catalogue and sources with $\rm{S/N}>3$ and flux density above $400~\rm{mJy}$ from Planck catalogues of Compact Sources and Galactic Cold Clumps at 353 GHz within a radius of 15$'$, then apodize the edges smoothly to unity with a Gaussian $\sigma$ of 0.559$^{\circ}$. This particular $\sigma$ ensures that the argument of the exponential is at $-5$ by a radius of 1.5$^{\circ}$ from each source. We normalize the resulting mask then compute $f_{\rm{sky}}^{\rm{eff}}$ from $f_{\rm{sky}}w_2^2/w_4$ (equation (17) in \citeauthor{hivon/etal/2002} \citeyear{hivon/etal/2002}), where $w_i$ is the $i$th moment of the mask, to find 0.62, 0.45, and 0.28 for our three masks. The main results presented in the following sections are not sensitive to the exact details of the apodization used here.\\

\section{Method}
\label{sec:method}

We use the nominal full sky pseudo-$C_\ell$ approach \citep{hivon/etal/2002, kogut/etal/2003, brown/etal/2005}, which accounts for incomplete sky coverage and beam smoothing to give unbiased estimates of the angular power spectrum. We confirmed the accuracy of our code\footnote{The code uses \texttt{HEALPix} libraries for spherical harmonic transforms and \texttt{SLATEC} Fortran subroutine, DRC3JJ.F9, for the computation of Wigner-3j symbols.} with an independent code from Kendrick Smith, and also reproduced the 353 GHz power spectra shown in Figure 2 in P14 up to small differences seemingly due to the differences in the assumed masks. The pure pseudo-$C_\ell$ estimator is not needed in this analysis since the power in E- and B-modes are similar for foregrounds and our masks retain large sky fractions.

We use only cross spectra in estimating the signal to avoid noise bias in our estimates. The power spectrum at 353 GHz is obtained from the cross spectra of the half-mission maps. For the cross spectra between WMAP and Planck, we use the WMAP single year maps and the Planck half-mission maps, then give the weighted mean for each cross-frequency spectra, accounting for the covariance among many cross spectra as we describe below. For all auto-\footnote{We use the term auto-DA spectra to refer to the cross spectra associated with one DA. The term auto power spectrum refers to the power spectrum of a map of signal plus noise. } and cross-DA spectra within WMAP, we use the nine single year maps to get 36 cross spectra for each DA, and 81 cross spectra for each cross-DA pair. Then we report the five single- and ten cross-frequency spectra for WMAP as the weighted mean of all spectra involving the relevant DAs. The W4 channel is omitted due to its failure of internal consistency tests \citep{jarosik/etal/2011}, which we independently confirmed. This channel had a large in-flight $1/f_{\rm{knee}}$ compared to those of the other DAs \citep{jarosik/etal/2003}. We also note that all maps are used at their provided resolution: $N_{\mathrm{side}}=512$ for WMAP and $N_{\mathrm{side}}=2048$ for Planck. 

Our goal is to parametrize the foreground emission, a non-Gaussian signal, with representative error bars. To this end we make a number of approximations in estimating the statistical uncertainty as described below. In particular, for WMAP we derive the statistical uncertainty from the distribution of the spectra formed with the nine single year maps, as each measures essentially the same sky (in one year, WMAP measures the full sky twice). A similar treatment is shown in \cite{hinshaw/etal/2009}, although it is for $\ell\leq10$ and for foreground-cleaned maps, which should be close to Gaussian. Detailed studies at large angular scales, say $\ell \lesssim 20$, need to use the pixel-based likelihood solution with the pixel-based noise covariance matrix (Appendix D of P07) or even better, simulations based on the time-ordered data. However, for our purposes, the pixel-based likelihood is not needed.

\begin{figure}
\centering
\epsscale{1.15} 
\plotone{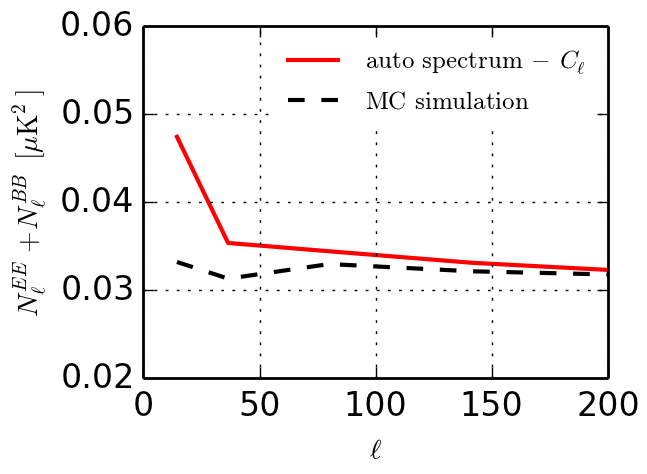}
\caption{The beam-deconvolved noise power spectra of the 9-year K-band maps are shown for the mask with $f_{\rm{sky}}^{\rm{eff}}$ of 0.62. The solid red line corresponds to $N_b^{EE} + N_b^{BB}$ from subtracting each $C_b$ estimated from the 36 cross spectra of the single year maps from each auto power spectrum of the 9-year maps. The black dashed line corresponds to $N_b^{EE} + N_b^{BB}$ obtained from the mean of each spectra from 1000 MC simulations of the noise maps. A similar quantity illustrating the large scale noise correlation is shown in Figure 16 of P07.}
\label{fig:Kband_N_b}
\end{figure}

To increase the signal to noise ratio, we compute the power spectra binned in multipole bands $b$ of width $\Delta\ell_b$ specified below. The auto power spectrum of a single map in bin $b$ gives $C_b+N_b$, where $C_b$ is the binned signal power spectrum, which includes both foregrounds and the CMB, and $N_b$ is the binned noise power spectrum. We can estimate $N_b$ from the auto power spectrum by subtracting $C_b$ estimated from the cross spectra. The estimate of $N_b$ here is then sensitive to any $1/\ell$-like component of the noise, since the particular realization of the noise contained in the data retains the large scale correlations from the 1/f noise in the time-ordered data. Alternatively, we can estimate $N_b$ from the Monte Carlo (MC) simulations of the noise sky using the provided noise maps (the diagonal element of the full noise covariance matrix). These simulations, without the full noise covariance matrix, do not capture the $1/\ell$ component of the noise, as shown in Figure~\ref{fig:Kband_N_b} for the 9-year K-band maps.

\newcolumntype{g}{>{\columncolor[gray]{0.85}}c}
\begin{table*}[ht]
\caption{Single and cross-frequency spectra $\chi^2 / \mathrm{dof}$}
\centering
\begin{adjustbox}{max width=\textwidth}
\begin{tabular}{| c | c | c | c | g | c | g | c | g | c | g | c | g | c | g |}
\hline
Channel & $\nu$ [GHz]\footnote{The effective frequency centers for the given foreground spectrum (see details in Section~\ref{sec:method}). The single-frequency spectra show the obtained physical frequency centers and each cross-frequency spectrum shows the geometric mean of the two frequencies for the pair.} & $\mathrm{dof}$\footnote{The degrees of freedom (dof) for the $\chi^2$ in each set of spectra, accounting for the removal of the mean.} & \multicolumn{4}{c |}{$\ell \in [6, 23]$} & \multicolumn{4}{c|}{$\ell \in [24, 49]$} &  \multicolumn{4}{c |}{$\ell \in [50, 110]$} \\
\hline
& & & \multicolumn{2}{c |}{EE} & \multicolumn{2}{c |}{BB} & \multicolumn{2}{c |}{EE} & \multicolumn{2}{c |}{BB} & \multicolumn{2}{c |}{EE} & \multicolumn{2}{c |}{BB} \\
\hline
K & 22.4 & 35 & \textbf{2.10}\footnote{The reduced $\chi^2$ for single and cross-frequency spectra in the $f_{\rm{sky}}^{\rm{eff}}$ of 0.62 region. Boldface denotes $>10\%$ difference between $\chi^2 / \mathrm{dof}$ from using the full covariance matrix and using just the diagonal elements.}  & \textbf{2.75}\footnote{Shaded columns are $\chi^2 / \mathrm{dof}$ after $\nu_b$ calibration from $g_b$. See Section~\ref{sec:method} for details.}  & \textbf{1.64} & \textbf{1.77} & \textbf{1.43} & \textbf{1.20} & \textbf{0.77} & \textbf{0.58} & 0.99 & 0.78 & 1.11 & 0.91\\
KKa & 27.1 & 80 & \textbf{6.24} & \textbf{7.29} & \textbf{4.79} & \textbf{5.09} & \textbf{1.52} & \textbf{1.13} & \textbf{1.39} & \textbf{0.98} & 1.09 & 0.74 & 2.02 & 1.43\\
KQ & 30.1 & 161 & \textbf{8.15} & \textbf{9.02} & \textbf{2.23} & \textbf{2.40} & \textbf{2.06} & \textbf{1.51} & \textbf{1.97} & \textbf{1.34} & 1.14 & 0.73 & 1.27 & 0.83\\
KV & 36.7 & 161 & \textbf{4.96} & \textbf{5.33} & \textbf{2.64} & \textbf{2.84} & \textbf{1.56} & \textbf{1.24} & \textbf{1.75} & \textbf{1.23} & 1.11 & 0.68 & 1.33 & 0.84\\
KW & 45.8 & 242 & \textbf{4.15} & \textbf{4.34} & \textbf{3.04} & \textbf{3.30} & \textbf{2.18} & \textbf{1.57} & \textbf{1.42} & \textbf{0.94} & 1.31 & 0.83 & 1.33 & 0.81\\
Ka & 32.7 & 35 & \textbf{1.31} & \textbf{1.40} & \textbf{1.91} & \textbf{1.60} & 0.72 & 0.55 & 1.32 & 0.94 & 0.80 & 0.65 & 1.20 & 1.00\\
KaQ & 36.3 & 161 & \textbf{2.07} & \textbf{1.85} & \textbf{1.30} & \textbf{0.99} & 1.40 & 1.02 & 1.28 & 0.88 & 1.45 & 1.04 & 1.28 & 0.99\\
KaV & 44.3 & 161 & \textbf{1.63} & \textbf{1.40} & \textbf{1.41} & \textbf{1.10} & 1.35 & 0.92 & 1.36 & 0.97 & 1.26 & 0.91 & 1.26 & 0.91\\
KaW & 55.2 & 242 & \textbf{1.69} & \textbf{1.45} & \textbf{1.53} & \textbf{1.14} & 1.42 & 1.06 & 1.45 & 1.02 & 1.21 & 0.83 & 1.37 & 0.95\\
Q & 40.3 & 152 & \textbf{2.01} & \textbf{1.40} & 1.82 & 1.24 & 1.39 & 0.96 & 1.56 & 1.08 & 1.47 & 1.13 & 1.29 & 1.01\\
QV & 49.3 & 323 & 1.89 & 1.27 & 1.64 & 1.11 & 1.40 & 0.96 & 1.52 & 1.01 & 1.09 & 0.83 & 1.42 & 1.08\\
QW & 61.4 & 485 & 1.74 & 1.16 & 2.02 & 1.27 & 1.51 & 1.04 & 1.36 & 0.89 & 1.38 & 1.04 & 1.26 & 0.94\\
V & 60.2 & 152 & 1.61 & 1.05 & 2.20 & 1.43 & 1.50 & 1.02 & 1.35 & 0.91 & 1.23 & 0.99 & 1.39 & 1.09\\
VW & 75.0 & 485 & 1.77 & 1.14 & 2.43 & 1.56 & 1.42 & 1.02 & 1.73 & 1.27 & 1.21 & 0.92 & 1.21 & 0.91\\
W & 93.4 & 350 & 1.63 & 1.05 & 2.95 & 1.93 & 1.48 & 1.01 & 1.52 & 1.08 & 1.28 & 0.99 & 1.09 & 0.85\\
\hline
K353 & 90.4 & 17 & \textbf{1.10} & \textbf{1.14} & \textbf{1.01} & \textbf{1.32} & \textbf{1.22} & \textbf{1.06} & \textbf{1.39} & \textbf{1.25} & \textbf{0.87} & \textbf{0.77} & \textbf{0.90} & \textbf{0.73}\\
Ka353 & 109.1 & 17 & \textbf{1.51} & \textbf{1.66} & \textbf{1.69} & \textbf{1.88} & \textbf{0.44} & \textbf{0.38} & 0.65 & 0.55 & \textbf{1.15} & \textbf{1.08} & 0.51 & 0.46\\
Q353 & 121.2 & 35 & \textbf{1.08} & \textbf{1.21} & 1.15 & 1.28 & \textbf{1.49} & \textbf{1.34} & \textbf{0.96} & \textbf{0.86} & \textbf{1.02} & \textbf{0.98} & 0.61 & 0.55\\
V353 & 148.1 & 35 & \textbf{1.36} & \textbf{1.52} & \textbf{1.08} & \textbf{1.15} & \textbf{0.88} & \textbf{0.81} & 1.40 & 1.27 & 0.64 & 0.59 & \textbf{0.85} & \textbf{0.75}\\
W353 & 184.4 & 53 & \textbf{1.67} & \textbf{1.79} & \textbf{0.98} & \textbf{1.08} & 1.27 & 1.15 & 1.52 & 1.41 & \textbf{1.09} & \textbf{1.06} & \textbf{0.91} & \textbf{0.84}\\
353 & 364.2 & 0 &  &  &  &  &  &  &  &  &  &  &  & \\
\hline
\end{tabular}
\end{adjustbox}
\label{tab:chi2}
\end{table*}

To determine the uncertainty for a set of cross spectra from many data splits, we need the covariance matrix that describes the correlations among the cross spectra with common data splits. The usual Gaussian approximation of the total uncertainty for an auto power spectrum in bin $b$ is given by
\begin{equation}
(\Delta C_b)^2 = \frac{2}{\nu_b}(C_b+N_b)^2,
\label{eq:auto_error}
\end{equation}
with $\nu_b$ defined as,
$$\nu_b = (2\ell_b + 1)\Delta\ell_b f_{\rm{sky}}^{\rm{eff}},$$
where $\ell_b$ is the central multipole in $b$, and $f_{\rm{sky}}^{\rm{eff}}$ is the effective fraction of sky as described in Section~\ref{sec:data}. The $C_b^2$ term is the sample variance. We generalize this expression for cross spectra uncertainties following the appendix in \cite{das/etal/2011}. The covariance matrix element of cross spectra for each bin $b$ is given by
\begin{equation}
\begin{split}
\Theta_{\mathrm{bb}}^{(\mathit{i}A \times \mathit{j}B);(\mathit{k}C \times \mathit{l}D)} = \frac{1}{\nu_b} 
 \Big[ & \Big \langle C_b^{(\mathit{i}A \times \mathit{k}C)}\Big \rangle \Big \langle C_b^{(\mathit{j}B \times \mathit{l}D)} \Big \rangle \\
+ & \Big \langle C_b^{(\mathit{i}A \times \mathit{l}D)}\Big \rangle \Big \langle C_b^{(\mathit{j}B \times \mathit{k}C)} \Big \rangle \Big],
\end{split}
\label{eq:cross_error}
\end{equation}
where different instrument channels and data splits are labeled with uppercase and lowercase roman letters respectively. Note that we recover equation~(\ref{eq:auto_error}) for the case of auto spectra, $\mathit{i} = \mathit{j} = \mathit{k} = \mathit{l}$ and $A=B=C=D$, since 
$\langle C_b^{(\mathit{i}A \times \mathit{i}A)}\rangle = C_b+N_b$. 

When evaluating the distribution of the cross spectra from multiple splits of data, there is no sample variance as each data split contains the same signal spectrum; however, there are cross terms $C_bN_b$ between signal and noise, different in each spectrum. We illustrate this by considering equation~(\ref{eq:cross_error}) for WMAP K-band for bin $b$, a 36$\times$36 covariance matrix (for the spectra with nine single year maps excluding the auto spectra). The variance of year 1$\times$2 is
\begin{equation}
\begin{split}
\Theta_{\mathrm{bb}}^{(\mathit{2}K \times \mathit{1}K);(\mathit{2}K \times \mathit{1}K)} = \frac{1}{\nu_b} 
\Big[ &\Big \langle C_b^{(\mathit{2}K \times \mathit{2}K)}\Big \rangle \Big \langle C_b^{(\mathit{1}K \times \mathit{1}K)} \Big \rangle \\
+ & \Big \langle C_b^{(\mathit{2}K \times \mathit{1}K)}\Big \rangle \Big \langle C_b^{(\mathit{2}K \times \mathit{1}K)} \Big \rangle \Big],
\end{split}
\end{equation}
which evaluates to 
$$\Theta_{\mathrm{bb}}^{(\mathit{2}K \times \mathit{1}K);(\mathit{2}K \times \mathit{1}K)} \propto \Big [N_b^{2K}N_b^{1K} + C_b^K(N_b^{2K} + N_b^{1K}) \Big ].$$
Similarly, the covariance of year 1$\times$2 with year 1$\times$3 is
$$\Theta_{\mathrm{bb}}^{(\mathit{2}K \times \mathit{1}K);(\mathit{3}K \times \mathit{1}K)} \propto \Big [C_b^K N_b^{1K} \Big ].$$
We can see that the matrix has the form where the diagonal elements include the $N\times N$ and the $C\times N$ terms, and the off diagonal elements have just the $C\times N$ terms or zeros. 

The effective number of modes $\nu_b$, parametrized by our estimate of $f_{\rm{sky}}^{\rm{eff}}$, needs to be calibrated with simulations to account for the spatial inhomogeneity and correlations in the noise and the complex mask. In particular, to calibrate $\nu_b$ at all angular scales one needs the pixel-based covariance matrix to simulate the noise maps and a template for the signal. With these, the distribution of the power spectra has the representative level of variance from the $C\times N$ and the $N\times N$ terms, but does not have sample variance. 
For the native resolution and our $\ell$ range, such a calculation is computationally intractable. 
Hence we instead use the uncertainties associated with each pixel (the diagonal elements of the full pixel-based covariance matrix) including the $QU$ correlations to generate the noise map and simulate the signal sky from $C_\ell$, a power law in $\ell$ fitted to the estimates of $C_b$ for each single- and cross-frequency spectra. Then we get the statistical variance of the power spectrum of the simulations, $(\Delta C_b^{\rm{MC}})^2$, by subtracting the sample variance (obtained from the signal-only simulations) from the variance in the signal plus noise simulations.
The $f_{\rm{sky}}^{\rm{eff}}$ calibration factor is given by the ratio of the statistical variance in the simulations to the analytic expression from equation~(\ref{eq:cross_error}),
$$g_b^{(A\times B)} = (\Delta C_b^{\rm{MC}})^2/\Theta_{\mathrm{bb}}^{(A \times B);(A \times B)}, $$
where the signal and noise power spectra in $\Theta_{\mathrm{bb}}^{(A \times B);(A \times B)}$ and $(\Delta C_b^{\rm{MC}})^2$ are from 2000 simulations. Note this method of obtaining $g_b$ takes into account the effects from the noise inhomogeneity and the $QU$ correlations but not the 1/$\ell$ noise component.

For one $\ell$-bin $b$, there are 20 covariance matrices\footnote{For all single- and cross-frequency spectra except 353 GHz.} of dimension $\rm{dof}$+1, where $\rm{dof}$ is given in the third column of Table~\ref{tab:chi2}. We construct each covariance with the estimate of $C_b$, initially from the mean of all spectra involving the relevant DAs, and the estimate of $N_b$ from the auto spectra, as described above (thus $N_b$ here contains the $1/\ell$ component of the noise).
Then with each auto-DA covariance matrix, we get the weighted mean of the cross spectra to estimate $C_b$, use this estimate to rebuild the covariance matrix, then iterate until convergence\footnote{Convergence is achieved with 3 iterations.}. This second step typically shifts $C_b$ by less than 0.1$\sigma$, though the shift could be as much as 2.8$\sigma$. The shift essentially shows the difference between the weighted mean and the initial arithmetic mean of the cross spectra. The final estimates of the auto-DA $C_b$ obtained this way are the signal input for the cross-frequency covariance matrices. When the WMAP channels are noise-dominated, the auto-DA $C_b$ can fluctuate negative. In such case, we simply use the auto spectra as the estimate of $N_b$ and also ignore the $C_b \times N_b$ terms when building the covariance matrices to avoid any non-physical noise representation in the data. This happens in 53 out of 162 auto-DA $C_b$ for all bins and masks\footnote{9 (DA) $\times$ 2 (EE, BB) $\times$ 3 ($\ell$-bins) $\times$ 3 (masks) = 162}, with 31 of them in the W-band and the rest in Q or V bands.

The bandpass mismatch within each band, say, between the WMAP channels Q1 and Q2, leads to a small amount of fluctuation among the single-frequency spectra and the cross-frequency spectra with the other channels. WMAP calibrates on the CMB. However, the low frequency channels are dominated by synchrotron emission and have different central frequencies. Thus, Q1 and Q2 have slightly different synchrotron levels. We have combined all of the cross spectra from Q1, Q2, and Q1$\times$Q2, then give the effective central frequency in Table~\ref{tab:chi2}. We do not scale the spectra from the different channels before combining them, and this affects our $\chi^2/\rm{dof}$ by $<1\%$. The frequency centers in Table~\ref{tab:chi2} were determined iteratively considering the spectral index of the signal in antenna temperature\footnote{We use all spectra in the first $\ell$-bin for EE and BB simultaneously to find the spectral index at each band, compute the frequency center corrections for the given indices, recompute the spectral indices at the new frequency centers, and iterate the process until convergence.}. For WMAP, we use the software on LAMBDA\footnote{\url{http://lambda.gsfc.nasa.gov/product/map/current/effective_freq.cfm}} to get the corrections, which are based on \cite{jarosik/etal/2003}. For Planck, we use the corrections in Table 4 of \cite{planck_ix/2013}. This ensures our estimates are the representative power at the given effective central frequencies.

We use the covariance matrices with no sample variance to evaluate the consistency of the measurements and the error bars. 
We show the reduced $\chi^2$ for all single- and cross-frequency spectra in Table~\ref{tab:chi2} for the mask with $f_{\mathrm{sky}}^{\mathrm{eff}}$ of 0.62 before and after calibrating $\nu_b$ with $g_b$. The cross-frequency spectra here combine all corresponding cross-DA spectra. The left hand column in each pair shows $\chi^2$ from before calibrating $\nu_b$.  After correcting $\nu_b$ for the number of modes that are lost, the scatter in the data at low $\ell$ is large even though the $C\times N$ terms and the $1/\ell$ component of the noise are accounted for. This is likely due to not using the full pixel-based covariance matrix. The departure from unity in the $\chi^2$/dof corresponds to an underestimate of the uncertainty up to a factor of $\approx3$ for WMAP and an overestimate up to a factor of $\approx1.5$ in WMAP $\times$ Planck. In general, though, we find that we can treat the signal as normally distributed for $\ell>23$. 

As a consistency check of the uncertainties, we compare our estimates from the data split maps (single year and half-mission maps) to those estimated with the full coadded maps (9-year and full-mission maps). First we subtract $C_b$ estimated from the data split maps from the auto spectra of the full coadded maps to get $N_b$. We then get the statistical uncertainty with equation~(\ref{eq:cross_error}). 
For the cross-frequency spectra, the agreement of the uncertainties is typically within a few percent with a deviation of $\sim30\%$ occurring at K$\times$353 for the first $\ell$-bin in EE for each mask. For the single-frequency spectra, our nominal uncertainties using the data split maps are greater than those estimated from the full coadded maps by typically $\sim5\%$, as the auto spectra are not included for the single-frequency spectra, while $N_b$ from the full coadded maps include the full data\footnote{For instance, out of the total 45 possible spectra within WMAP K-band, 9 auto spectra are not included. This amounts to a factor of $\sqrt{9/8}$ larger error bar for the case of noise domination, compared to that from using the full 9-year map.}.

For fitting to the model, described in the next section, we need the covariance matrix of the 21 spectra in each $\ell$-bin $b$. To form this we use the $N_b$ from the full coadded maps, apply the $g_b$ factors, then scale the components appropriately to account for not including the auto power spectrum in our estimates of $C_b$. 

\section{Model}
\label{sec:model}
We seek a simple model to characterize the polarized foreground emission and to serve as a guide to the degree of cleaning necessary to identify primordial or lensing B-modes at $\ell<150$. A number of studies have been done previously for modeling and cleaning of the foreground (e.g., B13, P14, \citeauthor{fuskeland/etal/2014} \citeyear{fuskeland/etal/2014},  \citeauthor{dunkley/etal/2009} \citeyear{dunkley/etal/2009}, \citeauthor{planck_x/2015} \citeyear{planck_x/2015}, and \citeauthor{watts/etal/2015} \citeyear{watts/etal/2015}), to which we add by basing our model on the measured set of all single- and cross-frequency spectra with Planck 353~GHz and WMAP, as a function of $\ell$-bin $b$. One advantage of this approach is that variations with angular scale (or spectral bins) may be identified.

At the current level of precision, synchrotron and dust emission may be modeled with the WMAP K-band and the Planck 353 GHz channels as
\begin{equation}
\begin{split}
M(\nu) &= S(\nu) + D(\nu) \\
&= \alpha_s(\nu /\nu_K)^{\beta_s}\mathbb{S} + \alpha_d {\cal E}_d(\nu,\beta_d)\mathbb{D}, \\
\end{split}
\end{equation}
where $M(\nu)$ is a Stokes $Q$ or $U$ map in antenna temperature at frequency $\nu$ in GHz, ${\cal E}_d(\nu,\beta_d)$ is the Planck dust model from equation~(\ref{eq:dust_model}), $\mathbb{S}$ and $\mathbb{D}$ are normalized synchrotron and dust spatial templates at frequencies $\nu_K$ and 353\,GHz, with amplitudes $\alpha_s$ and $\alpha_d$. When taking the cross spectra between maps  $M(\nu_1)$ and $M(\nu_2)$, we are effectively looking at the covariance between the spatial templates weighted by the amplitude and the frequency dependence terms,
\begin{equation}
\label{eq:final_model}
\begin{split}
\mathcal{D}^{\nu_1 \times \nu_2} =  \widetilde{\alpha}_s^2(\nu_1 & \nu_2/\nu_K^2)^{\beta_s} + \widetilde{\alpha}_d^2 {\cal E}_d(\nu_1, \beta_d){\cal E}_d(\nu_2, \beta_d) \\
+ \rho_{sd} \widetilde{\alpha}_s \widetilde{\alpha}_d ((\nu_1& /\nu_K)^{\beta_s} {\cal E}_d(\nu_2, \beta_d) + (\nu_2/\nu_K)^{\beta_s}{\cal E}_d(\nu_1, \beta_d)), \\
\end{split}
\end{equation}
where $\widetilde{\alpha}_s$ ($\widetilde{\alpha}_d$) is defined as $\alpha_s$ ($\alpha_d$) times the variance in $\mathbb{S}$ ($\mathbb{D}$), and 
$\rho_{sd}$ is defined as,
\begin{equation}
\rho_{sd} \equiv <\mathbb{S} \times \mathbb{D}> / \sqrt{<\mathbb{S} \times \mathbb{S}><\mathbb{D} \times \mathbb{D}>},
\label{eq:rho}
\end{equation}
the covariance between dust and synchrotron templates divided by the square root of the product of the variance of each (free of noise bias). However, when we fit to this model, $\rho_{sd}$ is a free parameter that we check with the definition.

As shown in the next section, we can fix $\beta_d=1.59$ (P15) and obtain good fits between all single- and cross-frequency spectra formed with the WMAP and the Planck 353 GHz maps. Thus we consider just four parameters ($\widetilde{\alpha}_s, \, \beta_s, \, \widetilde{\alpha}_d,   \, \rho_{sd}$) as our nominal model, which we denote model A. Alternatively, we can fix $\rho_{sd}$ from equation~(\ref{eq:rho}) using the WMAP K-band and the Planck 353 GHz maps as the synchrotron and dust templates, and use ($\widetilde{\alpha}_s, \, \beta_s, \, \widetilde{\alpha}_d,  \, \beta_d$) as the free parameters, which we denote model B. We do not need more than four parameters for the 21 single- and cross-frequency spectra in each $\ell$-bin.  

Before fitting to the model, we subtract lensed $\Lambda$CDM CMB power with the tensor to scalar ratio $r=0$ from EE and BB, and add the appropriate cosmic variance to the uncertainty.  We then convert these foreground-only spectra into antenna temperature units, and fit the resulting set of spectra for each mask to equation~(\ref{eq:final_model}), using the full covariance matrix for each bin (described in the end of Section~\ref{sec:method}). 

The model parameters and the uncertainties are obtained from two ways. First, we simulate the data 10000 times using the covariance matrix, then for each simulation the parameters are found by minimizing the $\chi^2$ for the fit. We then give the central values and uncertainties from the maximum and $1\sigma$ of the distributions for each parameter. The priors we adopt for this method are uniform and largely unbounded except where they are non-physical,
$$\widetilde{\alpha}_s \geq 0$$
$$\widetilde{\alpha}_d \geq 0$$
$$\beta_s \leq 0$$
$$\beta_d \geq 0$$
$$\left|\rho_{sd}\right|\leq1.$$
We find consistent parameters from the posterior probability distributions sampled with the Markov Chain Monte Carlo (MCMC) methods using the \texttt{EMCEE} code \citep{foreman-mackey/etal/2013} with a conservative uniform prior distribution of $-4 \leq \beta_s \leq 0$. 

\section{Results}
\label{sec:results}
\begin{figure}
\epsscale{1.15} 
\plotone{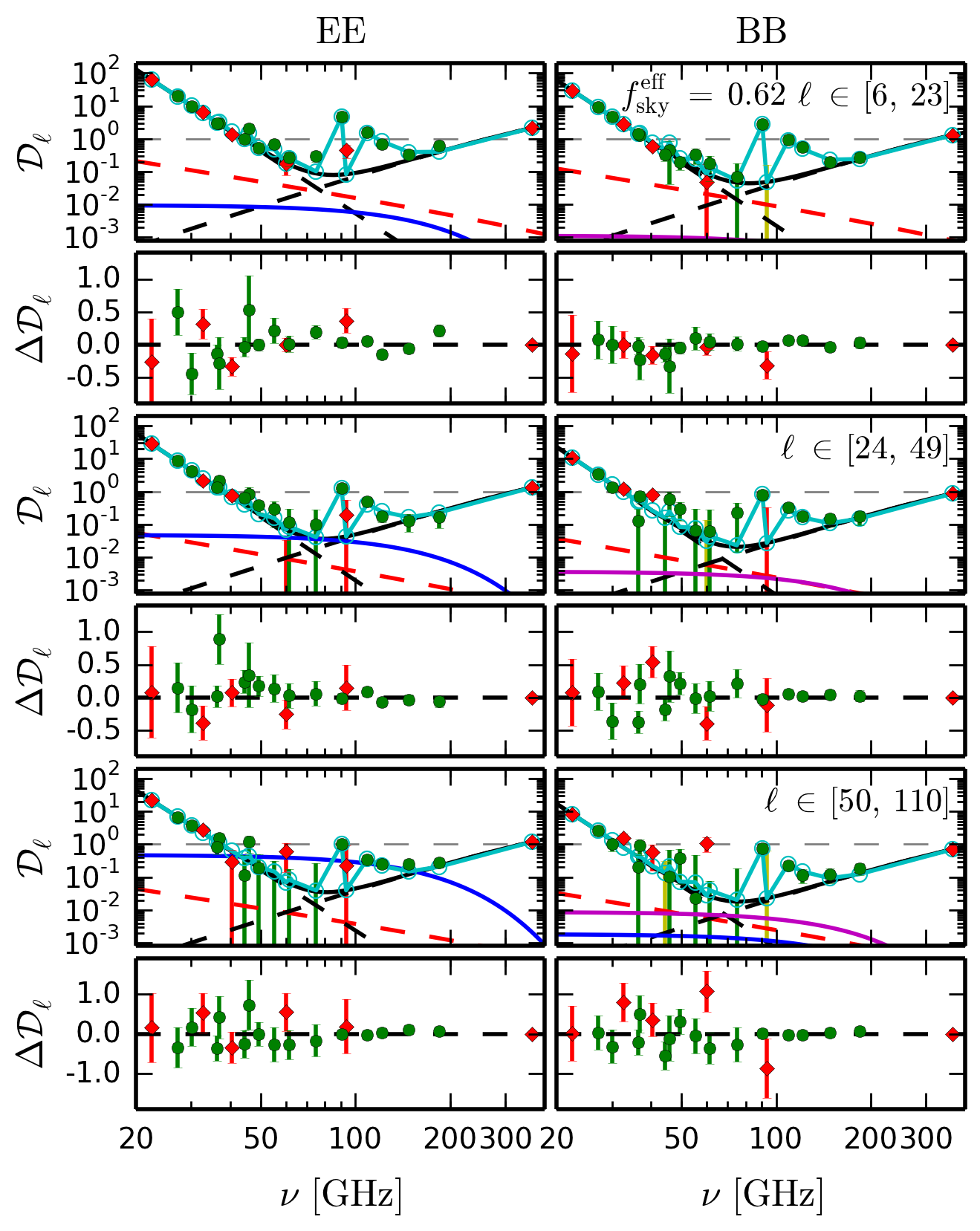}
\caption{The data and model for $f_{\rm{sky}}^{\rm{eff}}=0.62$. All y-axes are power in $\mu$K$^2$ in antenna temperature. The upper panel of each pair has a logarithmic y-axis whereas the lower panel is linear. Left and right columns correspond to EE and BB respectively. See Section~\ref{sec:results} for details.}
\label{fig:fsky70}
\end{figure}

\begin{figure}
\epsscale{1.15}
\plotone{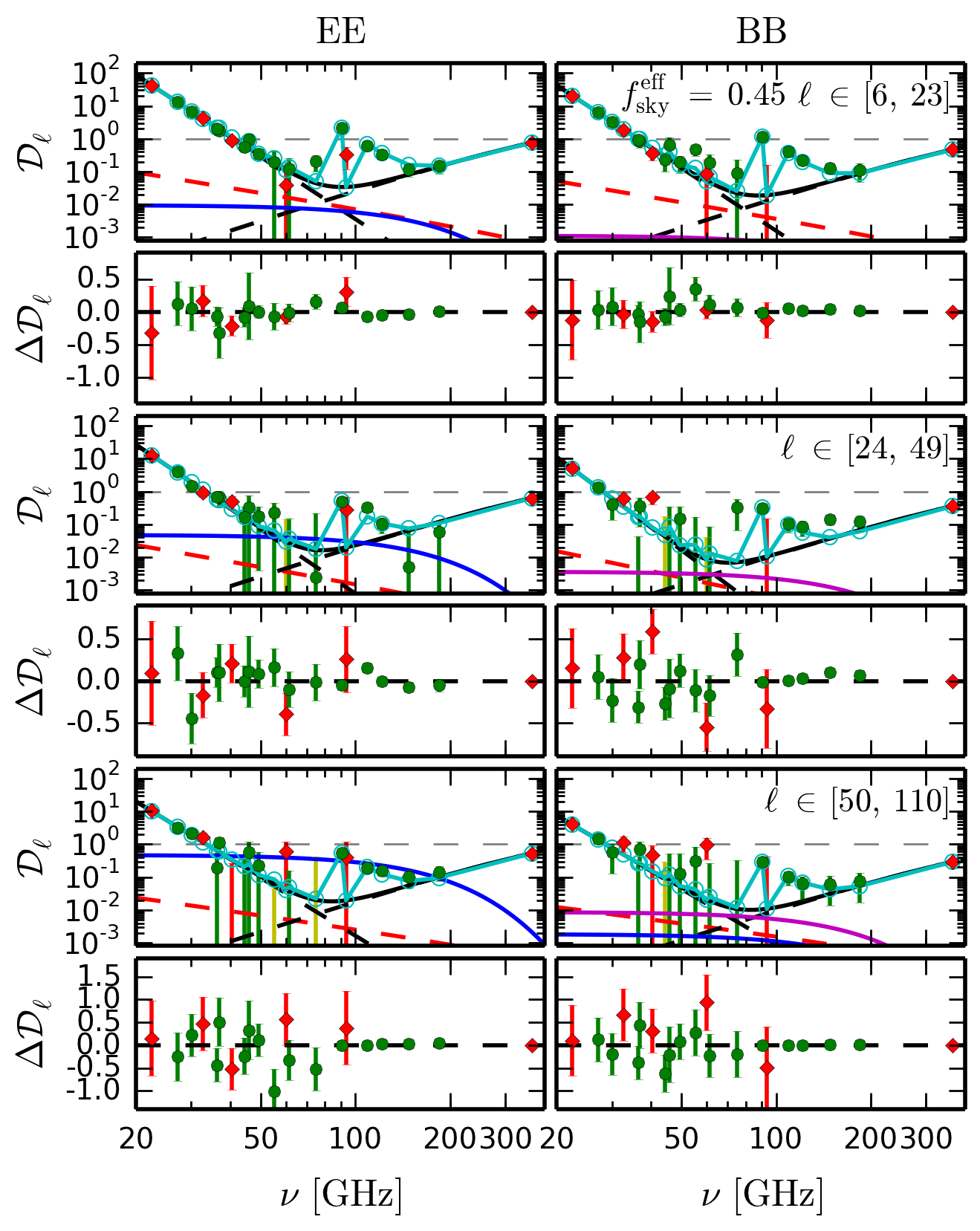}
\caption{The data and model for $f_{\rm{sky}}^{\rm{eff}}=0.45$ following the conventions of Figure~\ref{fig:fsky70}.} 
\label{fig:fsky50}
\end{figure}

\begin{figure}
\centering
\epsscale{1.15} 
\plotone{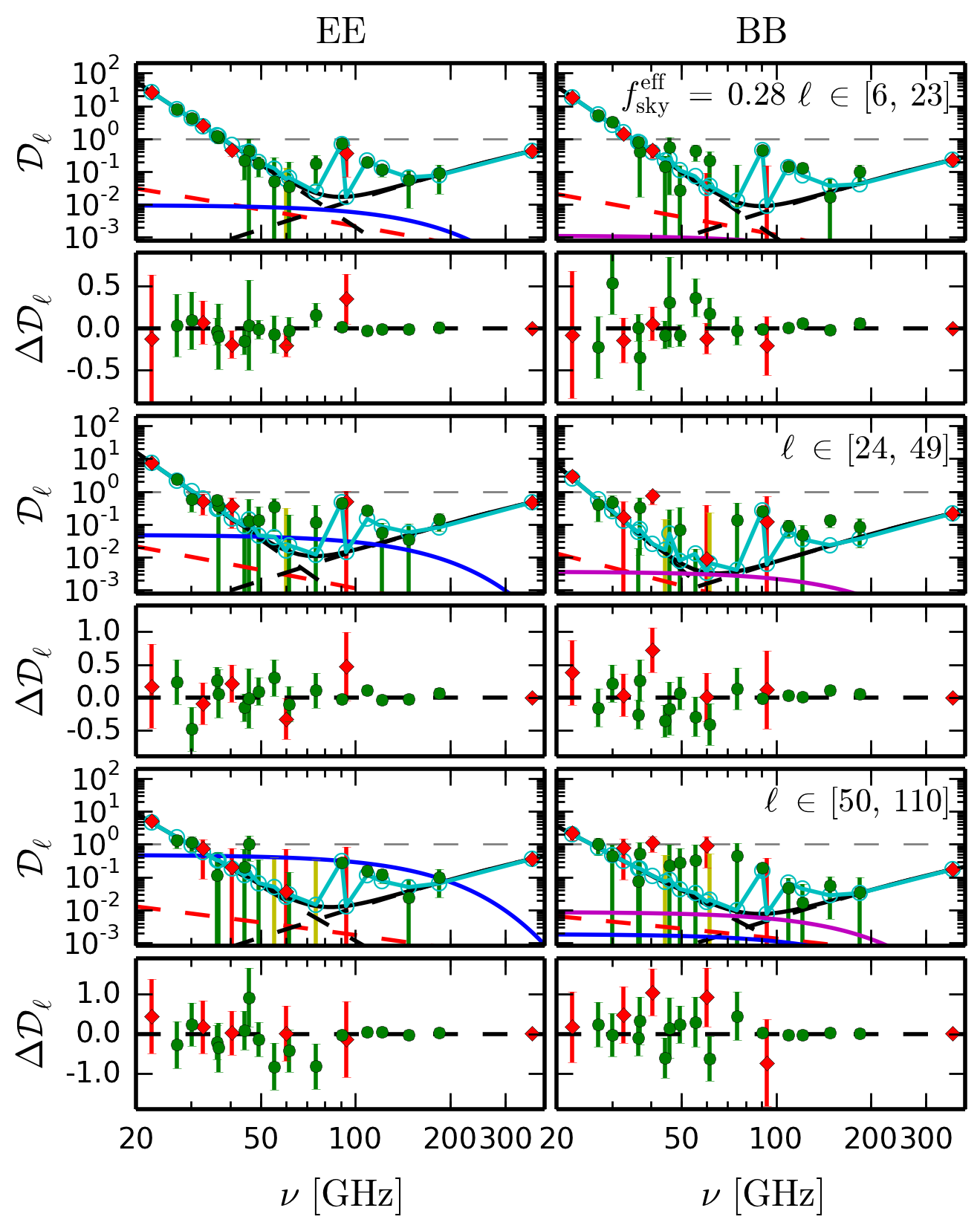}
\caption{The data and model for $f_{\rm{sky}}^{\rm{eff}}=0.28$ following the conventions of Figure~\ref{fig:fsky70}.}
\label{fig:fsky30}
\end{figure}

The power spectra in $\mathcal{D}_{\ell}$, where $\mathcal{D}_{\ell} \equiv \ell (\ell+1) C_\ell/(2\pi)$, and the corresponding fits to model A (from $\chi^2$ minimization) are shown in Figures~\ref{fig:fsky70},~\ref{fig:fsky50}, and~\ref{fig:fsky30} for the three masks. The spectra are CMB-subtracted and shown in antenna temperature units. The panels under each figure show the residuals on a linear scale.  The square root of the diagonal elements of the covariance matrices are shown as the error bars. The single-frequency spectra are shown in red diamonds, and the cross-frequency spectra are in green circles, plotted at the geometric mean of the two frequencies. The cross spectra do not necessarily represent the power at these plotted frequencies. The error bars that hit the x-axes pass through the zero line, and the yellow error bars denote two sigma error bars for when the spectrum estimate is negative.

In each panel, the blue line shows the $\Lambda$CDM CMB spectrum in antenna temperature. For BB, the magenta line shows the lensing B-mode signal plus an $r=0.1$ signal.  
The negatively-sloped black dashed line shows the synchrotron spectrum from $\widetilde{\alpha}_s^2$ and $\beta_s$. 
Similarly, the positively-sloped black dashed line shows the dust spectrum from $\widetilde{\alpha}_d^2$ and $\beta_d = 1.59$.
The dust-synchrotron correlated component is shown as a red dashed line. The expected total foreground power at any frequency is given by the smooth solid black curve (the sum of the three dashed lines), from equation~\ref{eq:final_model} for $\nu_1 = \nu_2$. Hence the cross-frequency spectra (green circles) are not generally expected to lie on this line. The fitted model is shown by the zigzag cyan line with cyan open circles, capturing both the single- and cross-frequency spectra. We emphasize the model plotted is valid only at the frequencies given in Table~\ref{tab:chi2} indicated by the open circles, and should not be interpolated between them. Lastly, the residuals to the best-fit model (the zigzag cyan line with cyan open circles) are shown in the panels below the spectra. A significant level of constraining power is seen in the cross spectra between Planck 353 GHz and WMAP (five green circles between 90-180 GHz).

\newcolumntype{f}{>{\columncolor[gray]{0.85}}r}
\begin{table*}[ht]
\caption{Results for model A}
\centering
\begin{adjustbox}{max width=\textwidth}
\begin{tabular}{| c | c | c | c | g | c | g | c | g | c | g | c | g | }
\hline
 $f_{\rm{sky}}^{\rm{eff}}$ &  $\ell$ & E/B & \multicolumn{2}{c |}{$\beta_s$} & \multicolumn{2}{c |}{$\rho_{sd}$} & \multicolumn{2}{c |}{$\widetilde{\alpha}_s^2$ ($\mu$K$^2$)\footnote{In antenna temperature units at the effective frequency center for K-band.}} & \multicolumn{2}{c |}{$\widetilde{\alpha}_d^2$($\mu$K$^2$)\footnote{In antenna temperature units at the effective frequency center for 353 GHz.}} & \multicolumn{2}{c |}{PTE \footnote{Note the results and the probability to exceed (PTE) values for the different masks are not independent as they include common regions. We assume the same $\mathrm{dof}=17$ for the two methods.}} \\
\hline
0.62 & 6-23 & EE & $-3.13^{+0.03}_{-0.03}$\footnote{Unshaded columns show the parameters found from $\chi^2$ minimization.}  & $-3.13^{+0.03}_{-0.03}$\footnote{Shaded columns show the parameters found from MCMC.}  & $0.41^{+0.01}_{-0.01}$  & $0.41^{+0.01}_{-0.01}$  & \multicolumn{1}{r|}{$62.4^{+0.6}_{-0.6}$}  & \multicolumn{1}{f|}{$62.4^{+0.6}_{-0.6}$}  & $2.11^{+0.01}_{-0.01}$  & $2.11^{+0.01}_{-0.01}$  & 0.02 \ & 0.02 \\
0.62 & 24-49 & EE & $-3.19^{+0.09}_{-0.09}$  & $-3.20^{+0.09}_{-0.09}$  & $0.20^{+0.01}_{-0.01}$  & $0.20^{+0.01}_{-0.01}$  & \multicolumn{1}{r|}{$28.2^{+0.7}_{-0.7}$}  & \multicolumn{1}{f|}{$28.2^{+0.7}_{-0.7}$}  & $1.31^{+0.01}_{-0.01}$  & $1.31^{+0.01}_{-0.01}$  & 0.46 \ & 0.46 \\
0.62 & 50-110 & EE & $-3.04^{+0.13}_{-0.14}$  & $-3.05^{+0.14}_{-0.14}$  & $0.20^{+0.01}_{-0.01}$  & $0.20^{+0.01}_{-0.01}$  & \multicolumn{1}{r|}{$21.8^{+0.8}_{-0.8}$}  & \multicolumn{1}{f|}{$21.8^{+0.8}_{-0.8}$}  & $1.17^{+0.01}_{-0.01}$  & $1.17^{+0.01}_{-0.01}$  & 0.74 \ & 0.74 \\
0.62 & 6-23 & BB & $-3.17^{+0.06}_{-0.06}$  & $-3.17^{+0.06}_{-0.06}$  & $0.45^{+0.01}_{-0.01}$  & $0.45^{+0.01}_{-0.01}$  & \multicolumn{1}{r|}{$29.4^{+0.6}_{-0.5}$}  & \multicolumn{1}{f|}{$29.4^{+0.6}_{-0.6}$}  & $1.29^{+0.01}_{-0.01}$  & $1.29^{+0.01}_{-0.01}$  & 0.32 \ & 0.32 \\
0.62 & 24-49 & BB & $-3.21^{+0.17}_{-0.17}$  & $-3.22^{+0.17}_{-0.18}$  & $0.27^{+0.02}_{-0.02}$  & $0.27^{+0.02}_{-0.02}$  & \multicolumn{1}{r|}{$10.9^{+0.5}_{-0.5}$}  & \multicolumn{1}{f|}{$10.8^{+0.5}_{-0.5}$}  & $0.86^{+0.01}_{-0.01}$  & $0.86^{+0.01}_{-0.01}$  & 0.20 \ & 0.20 \\
0.62 & 50-110 & BB & $-3.11^{+0.29}_{-0.30}$  & $-3.14^{+0.30}_{-0.29}$  & $0.30^{+0.03}_{-0.03}$  & $0.31^{+0.03}_{-0.03}$  & \multicolumn{1}{r|}{$8.4^{+0.7}_{-0.6}$}  & \multicolumn{1}{f|}{$8.3^{+0.7}_{-0.7}$}  & $0.70^{+0.01}_{-0.01}$  & $0.70^{+0.01}_{-0.01}$  & 0.41 \ & 0.42 \\
\hline
0.45 & 6-23 & EE & $-3.14^{+0.05}_{-0.05}$  & $-3.14^{+0.05}_{-0.05}$  & $0.37^{+0.01}_{-0.01}$  & $0.37^{+0.01}_{-0.01}$  & \multicolumn{1}{r|}{$42.6^{+0.7}_{-0.7}$}  & \multicolumn{1}{f|}{$42.5^{+0.7}_{-0.7}$}  & $0.76^{+0.01}_{-0.01}$  & $0.76^{+0.01}_{-0.01}$  & 0.16 \ & 0.16 \\
0.45 & 24-49 & EE & $-3.23^{+0.17}_{-0.19}$  & $-3.24^{+0.18}_{-0.18}$  & $0.18^{+0.02}_{-0.02}$  & $0.18^{+0.02}_{-0.02}$  & \multicolumn{1}{r|}{$12.7^{+0.6}_{-0.6}$}  & \multicolumn{1}{f|}{$12.7^{+0.6}_{-0.6}$}  & $0.63^{+0.01}_{-0.01}$  & $0.63^{+0.01}_{-0.01}$  & 0.19 \ & 0.19 \\
0.45 & 50-110 & EE & $-2.91^{+0.23}_{-0.26}$  & $-2.92^{+0.26}_{-0.26}$  & $0.23^{+0.03}_{-0.03}$  & $0.23^{+0.03}_{-0.03}$  & \multicolumn{1}{r|}{$10.3^{+0.8}_{-0.8}$}  & \multicolumn{1}{f|}{$10.3^{+0.8}_{-0.8}$}  & $0.53^{+0.01}_{-0.01}$  & $0.53^{+0.01}_{-0.01}$  & 0.60 \ & 0.60 \\
0.45 & 6-23 & BB & $-3.17^{+0.09}_{-0.09}$  & $-3.17^{+0.09}_{-0.09}$  & $0.37^{+0.01}_{-0.01}$  & $0.37^{+0.01}_{-0.01}$  & \multicolumn{1}{r|}{$20.6^{+0.6}_{-0.6}$}  & \multicolumn{1}{f|}{$20.6^{+0.6}_{-0.6}$}  & $0.47^{+0.01}_{-0.01}$  & $0.47^{+0.01}_{-0.01}$  & 0.53 \ & 0.53 \\
0.45 & 24-49 & BB & $-3.51^{+0.38}_{-0.45}$  & $-3.57^{+0.32}_{-0.32}$  & $0.24^{+0.04}_{-0.04}$  & $0.24^{+0.04}_{-0.04}$  & \multicolumn{1}{r|}{$5.0^{+0.5}_{-0.5}$}  & \multicolumn{1}{f|}{$5.0^{+0.5}_{-0.5}$}  & $0.36^{+0.01}_{-0.01}$  & $0.36^{+0.01}_{-0.01}$  & 0.08 \ & 0.08 \\
0.45 & 50-110 & BB & $-2.78^{+0.49}_{-0.55}$  & $-2.81^{+0.54}_{-0.51}$  & $0.26^{+0.05}_{-0.05}$  & $0.27^{+0.05}_{-0.05}$  & \multicolumn{1}{r|}{$4.1^{+0.7}_{-0.7}$}  & \multicolumn{1}{f|}{$4.0^{+0.7}_{-0.7}$}  & $0.29^{+0.01}_{-0.01}$  & $0.29^{+0.01}_{-0.01}$  & 0.91 \ & 0.90 \\
\hline
0.28 & 6-23 & EE & $-3.16^{+0.08}_{-0.08}$  & $-3.16^{+0.09}_{-0.09}$  & $0.21^{+0.02}_{-0.02}$  & $0.21^{+0.02}_{-0.02}$  & \multicolumn{1}{r|}{$26.3^{+0.7}_{-0.7}$}  & \multicolumn{1}{f|}{$26.3^{+0.7}_{-0.7}$}  & $0.42^{+0.01}_{-0.01}$  & $0.42^{+0.01}_{-0.01}$  & 0.85 \ & 0.85 \\
0.28 & 24-49 & EE & $-3.31^{+0.34}_{-0.38}$  & $-3.35^{+0.31}_{-0.33}$  & $0.25^{+0.04}_{-0.04}$  & $0.25^{+0.04}_{-0.03}$  & \multicolumn{1}{r|}{$7.5^{+0.7}_{-0.6}$}  & \multicolumn{1}{f|}{$7.4^{+0.6}_{-0.6}$}  & $0.47^{+0.01}_{-0.01}$  & $0.47^{+0.01}_{-0.01}$  & 0.48 \ & 0.48 \\
0.28 & 50-110 & EE & $-2.67^{+0.50}_{-0.55}$  & $-2.77^{+0.53}_{-0.55}$  & $0.23^{+0.05}_{-0.05}$  & $0.23^{+0.05}_{-0.05}$  & \multicolumn{1}{r|}{$4.8^{+0.9}_{-0.9}$}  & \multicolumn{1}{f|}{$4.6^{+0.9}_{-0.9}$}  & $0.36^{+0.02}_{-0.01}$  & $0.36^{+0.01}_{-0.02}$  & 0.93 \ & 0.92 \\
0.28 & 6-23 & BB & $-3.29^{+0.13}_{-0.13}$  & $-3.30^{+0.13}_{-0.14}$  & $0.22^{+0.02}_{-0.02}$  & $0.22^{+0.02}_{-0.02}$  & \multicolumn{1}{r|}{$18.6^{+0.7}_{-0.7}$}  & \multicolumn{1}{f|}{$18.5^{+0.7}_{-0.7}$}  & $0.23^{+0.01}_{-0.01}$  & $0.23^{+0.01}_{-0.01}$  & 0.65 \ & 0.65 \\
0.28 & 24-49 & BB & $-3.40^{+1.26}_{-0.90}$  & $-4.00^{+1.49}_{-0.00}$  & $0.35^{+0.07}_{-0.07}$  & $0.36^{+0.07}_{-0.07}$  & \multicolumn{1}{r|}{$2.8^{+0.5}_{-0.5}$}  & \multicolumn{1}{f|}{$2.5^{+0.5}_{-0.5}$}  & $0.22^{+0.01}_{-0.01}$  & $0.22^{+0.01}_{-0.01}$  & 0.32 \ & 0.31 \\
0.28 & 50-110 & BB & $-2.25^{+0.92}_{-1.27}$  & $-2.50^{+0.89}_{-0.79}$  & $0.25^{+0.11}_{-0.10}$  & $0.27^{+0.11}_{-0.10}$  & \multicolumn{1}{r|}{$2.3^{+0.8}_{-0.8}$}  & \multicolumn{1}{f|}{$2.0^{+0.8}_{-0.8}$}  & $0.17^{+0.01}_{-0.01}$  & $0.17^{+0.01}_{-0.01}$  & 0.86 \ & 0.85 \\
\hline

\end{tabular}
\end{adjustbox}
\label{tab:result}
\end{table*}

\begin{figure}[t!]
\centering
\epsscale{1.15}
\plotone{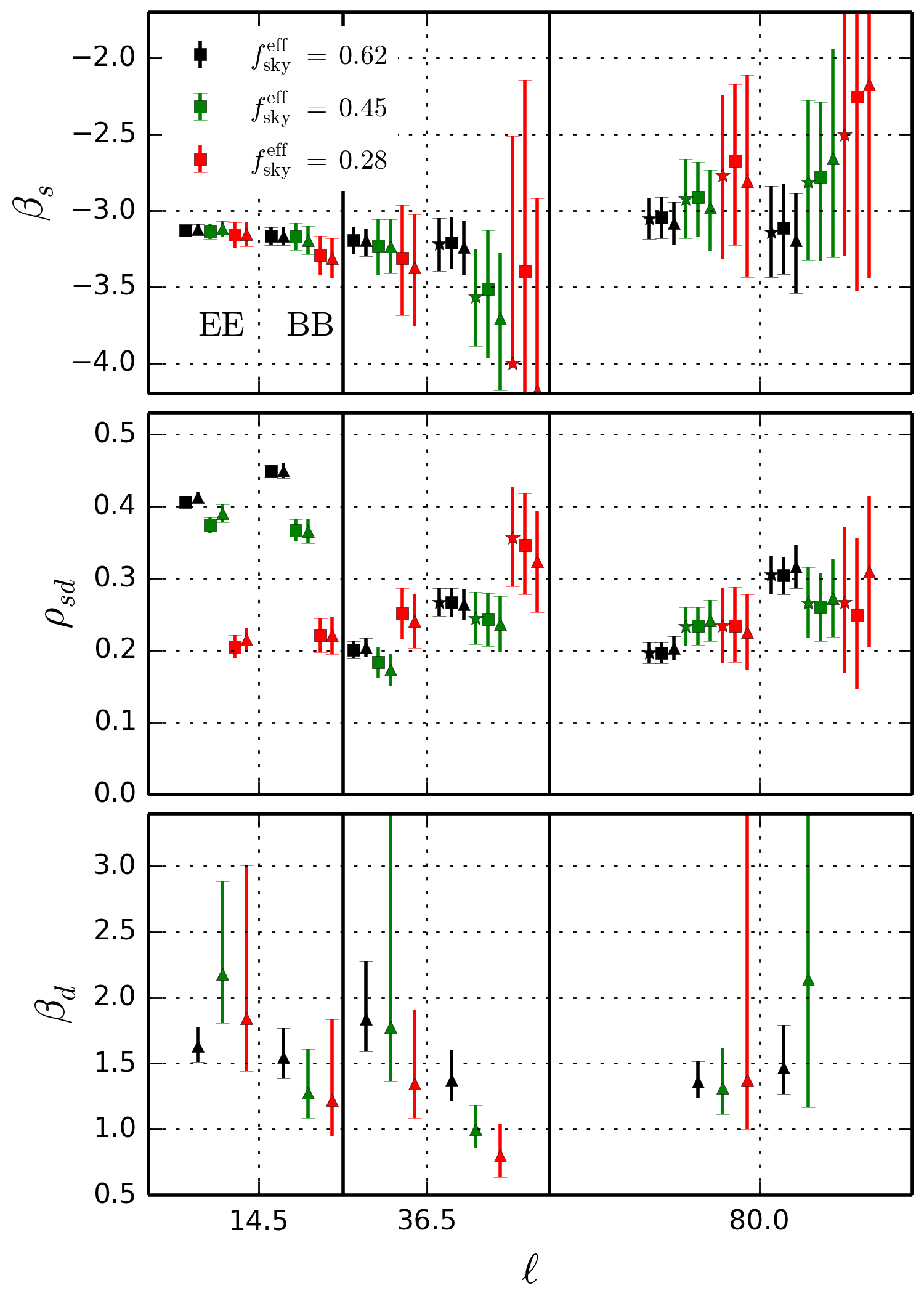}
\caption{The three panels show $\beta_s$, $\rho_{sd}$, and $\beta_d$ fitted for the two models for the three masks in the three $\ell$-bins. The results for each $\ell$-bin are separated by the solid black vertical lines. For each $\ell$-bin, to the left (right) of each dotted black line are the parameters for EE (BB), where different colors represent the results for the different masks. In each groups of points, the squares (triangles) show the results from model A (model B) from $\chi^2$ minimization. We show the parameters for model A from MCMC for the lower signal to noise ratio regime in stars. In the $f_{\rm{sky}}^{\rm{eff}}$ of 0.28 region, $\beta_s$ in the second $\ell$-bin is a 95$\%$ confidence limit from the prior bound at $-4$. For model A, $\beta_d$ is fixed, and thus not shown in the bottom panel.}
\label{fig:fitted_param}
\end{figure}

\begin{figure}
\centering
\epsscale{1.2}
\plotone{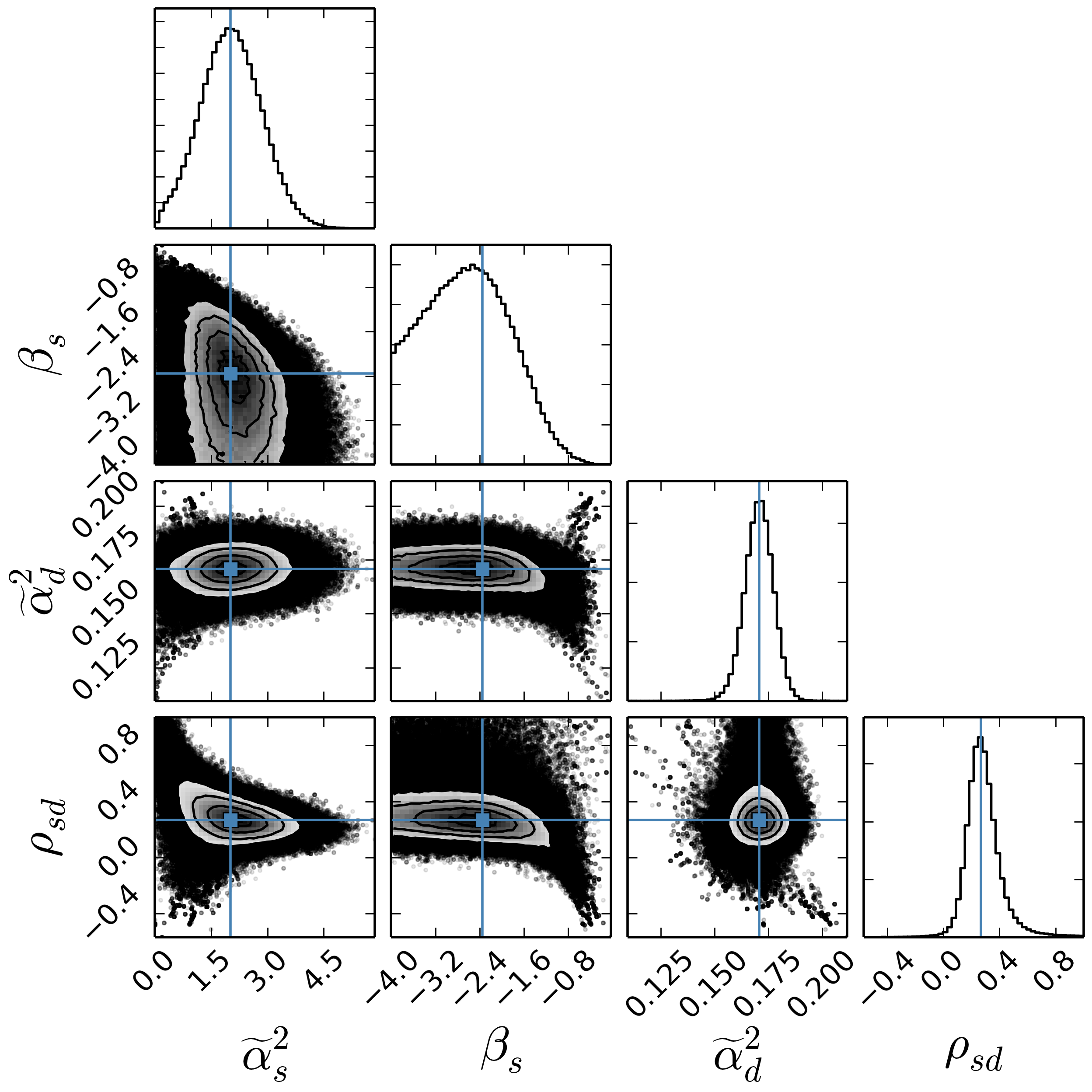}
\caption{The parameter probability distributions for $50 \leq \ell \leq 110$ in the $f_{\rm{sky}}^{\rm{eff}}$ of 0.28 region sampled with the MCMC methods. The panels on the diagonal show the marginalized distribution for each parameter. The parameters at the maxima of the probability distributions are shown in blue lines. The off diagonal panels show the joint, marginalized distributions, with the contours at 0.5, 1.0, 1.5, and 2.0$\sigma$. Note $\beta_s$ is cut off by the prior distribution. Figure made with \cite{foreman-mackey/etal/2014}.}
\label{fig:corner}
\end{figure}

The parameters of the best fit model found from $\chi^2$ minimization and MCMC are given in Table~\ref{tab:result}. The parameters obtained the two ways are consistent. Table~\ref{tab:result} shows that this simple model is a good fit to the data. Figure~\ref{fig:fitted_param} shows the fitted values of $\beta_s$, $\rho_{sd}$, and $\beta_d$. The top panel suggests that $\beta_s$ becomes more positive at finer angular scales although the large uncertainty precludes drawing a definitive conclusion. However, $\beta_s=-3.0$ does not appear to be less likely than $\beta_s=-3.3$, especially for $50 \leq \ell \leq 110$.  The center panel shows $\rho_{sd}$ from model A and also obtained from the templates (equation~(\ref{eq:rho})) for EE and BB, showing consistency between the two methods. We find $\rho_{sd} \approx 0.4$ for $6 \leq \ell \leq 23$ in the masks with $f_{\rm{sky}}^{\rm{eff}}$ of 0.62 and 0.45, then it drops to a constant value $\approx 0.2$. While the particular region we consider for $f_{\rm{sky}}^{\rm{eff}}$ of 0.28 shows smaller $\rho_{sd}$ in the first $\ell$-bin than those in the other two masks, we find a region with similar $f_{\rm{sky}}^{\rm{eff}}$ based on the dust intensity with Planck 857 GHz map also shows $\rho_{sd} \approx 0.4$. Note the largest angular scales not presented here show a larger level of correlation with $\rho_{sd} > 0.5$. The bottom panel shows the results for model B, with $\beta_d$ freed and $\rho_{sd}$ fixed from equation~(\ref{eq:rho}). We find consistent indices with those from P15. 

\begin{figure*}
\epsscale{1.2} 
\plotone{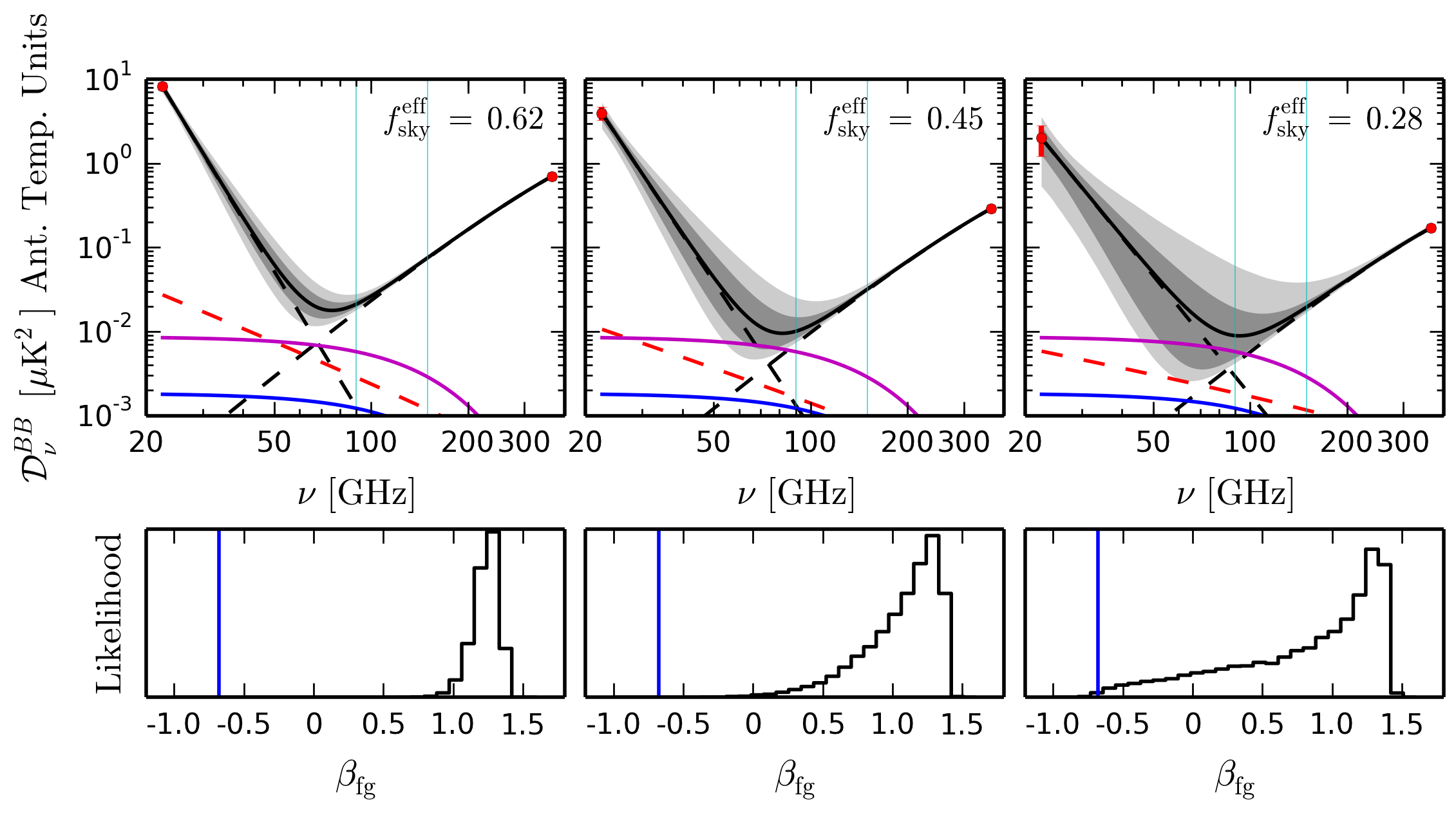}
\caption{Top panels show the best fit foreground model and the uncertainty from MCMC for BB in $50 \leq \ell \leq 110$ for the three masks, similar to the bottom right panels in Figures~\ref{fig:fsky70},~\ref{fig:fsky50}, and~\ref{fig:fsky30}. The model lines follow the same convention as before. Note a difference of $0.5\sigma$ in $\beta_s$ for the $f_{\rm{sky}}^{\rm{eff}}$ of 0.28 region can lead to a factor of $\sim4$ times larger level of synchrotron power at 100 GHz. The bottom panels show the likelihood of the foreground spectrum index between 90 and 150 GHz, along with that of the CMB spectrum at $\approx-0.7$ in blue. See Section~\ref{sec:results} for details.}
\label{fig:bb_bin2}
\end{figure*}

The $50 \leq \ell \leq 110$ region is the easiest place to search for primordial B-modes from the ground. We show the marginalized distributions on the parameters for $f_{\rm{sky}}^{\rm{eff}}$ of 0.28 in this $\ell$ range in Figure~\ref{fig:corner}. The current uncertainty on $\beta_s$ for such $f_{\rm{sky}}^{\rm{eff}}$ is large for this $\ell$ range, and this has a profound impact on the level of synchrotron emission near 100 GHz. A difference of $0.5\sigma$ in $\beta_s$ can lead to a factor of $\sim4$ times larger synchrotron power at 100 GHz. Figure~\ref{fig:bb_bin2} shows the best fit spectrum to model A (from MCMC) for the three different sky cuts. Although the fits to the model were obtained with $\beta_d$ fixed, we have also included the appropriate uncertainty on $\beta_d$ for the given $f_{\rm{sky}}^{\rm{eff}}$ (P14) in the shown uncertainty bands.
The dark (light) gray bands correspond to 1$\sigma$ (2$\sigma$) spread at each frequency in the individual fits from the accepted sets of parameters, illustrating the current uncertainty on polarized foregrounds in these regions. Future studies including the rest of the Planck polarization data will shrink these error bands. The red points in each panel show the best fit synchrotron and dust power at the reference frequencies\footnote{The frequency centers given for K-band and 353 GHz in Table~\ref{tab:chi2}.}.
The minimum frequency in the given $\ell$ range for the three masks is $\approx75$ GHz, similar to that for the temperature anisotropy (B13;  \citeauthor{planck_x/2015} \citeyear{planck_x/2015}).
For $f_{\rm{sky}}^{\rm{eff}}$ of 0.62 and 0.45 the ratio of synchrotron to dust emission at 90 GHz in amplitude is $0.3_{-0.1}^{+0.1}$ and $0.3_{-0.2}^{+0.3}$ respectively. In the $f_{\rm{sky}}^{\rm{eff}}$ of 0.28 mask, the uncertainty on the synchrotron emission is large. The ratio of synchrotron to dust amplitudes at 90 GHz is $<5$ at 95$\%$ confidence, and the ratio of the correlated component to the dust amplitudes is $0.2_{-0.1}^{+0.1}$.

The cyan lines in the top panels mark 90 and 150 GHz, typical frequencies for ground based observations. The index of the total foreground spectrum between these two frequencies, $\beta_{\rm{fg}}$, is shown in the bottom panels for each mask. Blue vertical lines at $\beta_{\rm{fg}}\approx-0.7$ indicate the index of the CMB spectrum between 90 and 150 GHz. 

Lastly, we examine the model in the region observed by the BICEP2 and Keck experiments \citep{bicepkeck/2015} which is cleaner in dust in the $50 \leq \ell \leq 110$ range than our $f_{\rm{sky}}^{\rm{eff}}=0.28$ region. We define the region by setting the pixels inside the published field outline\footnote{\url{http://bicepkeck.org/B2_3yr_373sqdeg_field_20140509.txt}} to 1 and the outside to $-0.01$ at  $N_{\mathrm{side}}=512$, smoothing with 6$^{\circ}$ Gaussian FWHM, then setting negative pixels to zero. The $f_{\rm{sky}}^{\rm{eff}}$ of the mask is 0.0088. Before computing the power spectra in the region, all maps are high pass filtered in harmonic space with an exponential function smoothly going from zero to one from $\ell$ of 2 to 21. The noise in WMAP is too large to make a significant measurement of $\widetilde{\alpha}_s$ in a single $\ell$-bin. Therefore we fix $\rho_{sd}=0.2$, $\beta_d=1.59$, and impose a Gaussian prior on $\beta_s=-3.0\pm0.3$ based on the results from the larger sky cuts, then fit the model using MCMC simultaneously in 3 even $\ell$ bins between 21 and 125 using $\mathcal{D}_\ell^{BB} = A_\mathrm{x} (\ell/80)^{-0.42}$ (P14) to get the synchrotron and dust power at $\ell=80$ ($A_s$ and $A_d$). The power law scalings is consistent with what we find in $\widetilde{\alpha}_s^2$ for $24 \leq \ell \leq 171$ in the larger masks. From the marginalized distributions we find $A_s < 2~\mu{\mathrm K}^2$ at 95$\%$ confidence and $A_d= 0.033\pm0.008~\mu{\mathrm K}^2$, in antenna temperature units at the given frequency centers. As it is based on: 1) $\beta_s$ and $\rho_{sd}$ from different regions of sky, 2) synchrotron and dust power from wide angular scales, and 3) different spatial weights than that used by the BICEP2 and Keck experiments, it is more of an expectation rather than a prediction. Note the dust level is $\sim$2 times smaller in amplitude than our $f_{\rm{sky}}^{\rm{eff}}=0.28$ region, but it is unclear how the synchrotron level scales due to the noise in WMAP.

\section{Discussion and Conclusions}
\label{sec:conc}

We have shown a basic polarized foreground model consisting of synchrotron, dust, and their correlation works for all possible single- and cross-frequency spectra formed with the WMAP9 and Planck 353 GHz data. The model needs four parameters: 1) synchrotron amplitude, 2) dust amplitude, 3) synchrotron spectral index, 4) synchrotron-dust correlation coefficient as a free parameter with the dust spectrum fixed with the Planck dust model, or the dust spectral index as a free parameter with the synchrotron-dust correlation coefficient fixed from the definition (equation~(\ref{eq:rho})) using synchrotron and dust templates.

We can draw a number of conclusions from the model.
1) The foreground minimum for BB in $50 \leq \ell \leq 110$ for the three regions we consider is $\approx$75 GHz, consistent with other studies \cite{bennett/etal/2013, planck_x/2015}.
2) There are three primary atmospheric windows for observing the CMB at 45, 90, and 150 GHz. For the cleanest 50$\%$ of sky in polarized dust emission, the ratios of the synchrotron to dust amplitudes at the three frequencies are approximately 9.0, 0.3, and 0.02 respectively.
3) In our model, any change in slope of the synchrotron index in frequency is explained as a cross correlation with dust. Thus, in each $\ell$-bin, one index describes the synchrotron, although it possibly flattens as one moves to smaller angular scales as shown in the top panel of Figure~\ref{fig:fitted_param}.  
4) The EE/BB ratio of power is approximately two although there are large variations depending on the sky cut and $\ell$ range.
5) When averaged over the cleanest $30\%$ of the sky in polarized dust, the total foreground emission power at the minimum in $50 \leq \ell \leq 110$ is $\approx0.01~\mu$K$^2$, corresponding to $r\approx0.15$. 

While the model is an excellent fit to the data, we note that foreground emission is subtle. At this point, there is no need for an additional component but that is simply due to the large uncertainties. One expects the synchrotron and dust spectral indices to vary spatially and there also may be magnetic dust emission \citep{draine/lazarian/1999}. The current data on polarized synchrotron emission are not constraining enough for an accurate determination of the precise levels in the cleanest region of sky. Instruments are being built for better measurements of the synchrotron emission (e.g., CLASS \citep{essinger-hileman/etal/2014}). 

\acknowledgments 
This work was supported by the U.S. National Science Foundation through PHY-1214379. We thank Kendrick Smith for sharing the pseudo-$C_\ell$ code, and Akito Kusaka for discussions of the method. We thank Norm Jarosik, Mike Nolta, and Ana-Roxana Pop for general discussions of the analysis, and Jo Dunkley, Raphael Flauger, Mark Halpern, and Eiichiro Komatsu for helpful comments that strengthened and clarified this paper. Lastly, we thank an anonymous referee for many useful comments, which further clarified the paper.

\bibliographystyle{yahapj} 
\bibliography{sync_dust_arxiv}
  
\end{document}